\documentclass[pdflatex,sn-mathphys-num]{sn-jnl}

\usepackage{graphicx}%
\usepackage{multirow}%
\usepackage{amsmath,amssymb,amsfonts}%
\usepackage{amsthm}%
\usepackage{mathrsfs}%
\usepackage[title]{appendix}%
\usepackage{xcolor}%
\usepackage{textcomp}%
\usepackage{manyfoot}%
\usepackage{booktabs}%
\usepackage{algorithm}%
\usepackage{algorithmicx}%
\usepackage{algpseudocode}%
\usepackage{listings}%
\usepackage{booktabs}
\usepackage{url}

\begin{document}

\title[Beyond Compliance: A User-Autonomy Framework for Inclusive and Customizable Web Accessibility]{Beyond Compliance: A User-Autonomy Framework for Inclusive and Customizable Web Accessibility}


\author*{\fnm{Lalitha} \sur{A R}}\email{24f2006078@ds.study.iitm.ac.in}

\affil{\orgdiv{Department of Data Science}, \orgname{Indian Institute Of Technology Madras}, \orgaddress{\postcode{600036}, \state{Tamil Nadu}, \country{India}}}

\abstract{
This paper proposes a shift from compliance-centered web accessibility to a care-driven model that prioritizes user autonomy, using neurodivergent users as a catalyst case for broader personalization needs. While accessibility standards offer a flexible framework, they are often interpreted and implemented as static compliance checklists, our approach reframes it as a flexible, user-centered process. We introduce a customizable \textit{Comfort Mode} framework that allows users to adapt interface settings, such as contrast, typography, motion, and scaling, according to their individual needs, while retaining the brand’s core visual identity. Grounded in psychological and cognitive accessibility principles, our design supports personalization without sacrificing creative freedom. We present both minimal and advanced implementation models with mock-ups, demonstrating how inclusive design can be seamlessly integrated at minimal cost. This approach aims to broaden digital inclusivity by offering autonomy to those who require it, without imposing changes on those who do not. The proposed system is adaptable, scalable, and suitable for a wide range of users and brands, offering a new paradigm where user autonomy, aesthetic integrity, and accessibility converge not through compromise, but through choice.
}

\keywords{Web accessibility, Inclusive design, User autonomy, Neurodiversity, Customization, Cognitive accessibility}

\maketitle

\section{Introduction}

In an increasingly digital-first world, the aspiration of universal access often encounters the complex reality of diverse human needs. While existing digital accessibility efforts provide a crucial foundation, they frequently struggle to accommodate the rich variability of human cognition and perception. This is particularly evident when considering neurodivergent individuals, whose unique sensory and cognitive processing styles powerfully illustrate why a 'one-size-fits-all' approach is insufficient. Instead, truly inclusive systems must foster personalization and user agency to bridge this gap.

Despite the maturation of accessibility guidelines such as the Web Content Accessibility Guidelines (WCAG), real-world implementation of comprehensive accessibility remains dismally low. An automated large-scale study of nearly three million web pages reported over 86 million accessibility errors, with fewer than 1\% of pages being error-free \cite{martins2024}. Similarly, the 2025 WebAIM Million study found over 50 million distinct accessibility errors across one million home pages, an average of 51 errors per page—suggesting that users with disabilities would encounter a barrier in roughly 1 out of every 24 elements \cite{webaim2025}. Recurring failures—such as poor contrast, missing alternative text, improperly labeled interface components, and misuse of ARIA—disproportionately affect users with cognitive and sensory processing differences who are already marginalized in digital environments.

This paper argues for a reframing of accessibility not as mere compliance, but as a form of care, one that honors user autonomy and supports interface adaptation. We use ‘care’ not simply as a metaphor, but in the ethical sense: as design attentive to lived context, individual dignity, and relational needs, rather than compliance with abstract norms \cite{Tronto1993}. Drawing a parallel from medicine \cite{medicineparallel}, we build on the distinction between "compliance" and "adherence" where the latter emphasizes individual agency over imposed norms. In our case, enabling users to customize how they engage with a web interface—through a simple but extensible \textit{Comfort Mode} toggle—offers a scalable pathway to inclusion. We use the term Comfort Mode to describe a user-controlled accessibility layer designed to enhance sensory and cognitive ease. It reframes accessibility not as remediation but as optional comfort—on the user’s terms.

As AI-driven interfaces and adaptive UI components become more common, anchoring personalization in ethical, user-centered frameworks is increasingly critical. Our approach offers such a foundation.

We propose a theoretical and practical framework for implementing such personalization. It preserves brand identity while allowing users who wish to customize their experience to do so, addressing shared needs across diverse user groups. We show how this functionality can be implemented using lightweight HTML, CSS, and JavaScript techniques, making it feasible for developers at any scale. In doing so, we advocate for a new standard of inclusive design—one where user dignity, design creativity, and accessibility coexist.

\section{Accessibility Is a Spectrum, Not a Checkbox}
Digital accessibility is often presented as a binary: a system is either compliant or non-compliant, accessible or inaccessible. However, this framing obscures the complex reality of user needs. Accessibility exists on a spectrum, not just across permanent disabilities but also along cognitive, sensory, and situational axes. For example, a user with dyslexia, a user with ADHD, and a user who is tired or overwhelmed may all benefit from different adjustments in typography, spacing, or motion—yet none of these needs fit neatly into binary criteria.

Research has emphasized the significance of recognizing cognitive accessibility as fluid and individualized. The World Wide Web Consortium’s (W3C) Cognitive and Learning Disabilities Accessibility Task Force explicitly notes that personalization is critical for supporting cognitive diversity, advocating for adaptable content, navigation, and presentation \cite{bib6}. They argue that one-size-fits-all accessibility fails to accommodate users who experience fluctuating attention, memory challenges, executive function issues, or processing delays.

Moreover, situational impairments—such as using a mobile device in direct sunlight, or multitasking while caring for a child can create temporary but significant access barriers. This underscores the value of designs that respond to a user’s current context, not just their fixed characteristics. Research shows that such contexts (e.g., bright outdoor lighting or motion while walking) can severely degrade usability unless interfaces adapt. For example, by adjusting visual contrast, simplifying layouts, or supporting gesture-based interactions, adaptive and universal design strategies are most effective in addressing these challenges \cite{sears2003, tigwell2018, steinfeld2019}.

Personalization, then, is not simply a feature—it is a core accessibility mechanism. Research consistently shows that adaptive interfaces—such as simplified text, glossaries, layout adjustments, and content filtering—can significantly improve usability for individuals with diverse cognitive needs, including but not limited to people with cognitive impairments and neurodivergent users~\cite{moreno2023simplification,hoehl2016web,waggoner2021inclusive}. Rather than enforcing a one-size-fits-all approach, inclusive systems empower users to choose the conditions under which they can most effectively process and engage with content.

This reframing aligns with the principle of “accessibility as user care.” It recognizes that good design accounts for variation over time, task, and individual context. By shifting from universal compliance to contextual adaptation, we move toward accessibility practices that prioritize user dignity, comfort, and autonomy.

\section{Divergent Approaches to Accessibility in Practise}
As previously discussed, Web accessibility is frequently treated as a binary outcome—either a site conforms to technical standards or it does not. This compliance-led mindset often stems from legal pressure, time constraints, or a desire to meet audit checkboxes. In practice, this has led many teams to frame accessibility as an afterthought or a risk-mitigation measure, rather than a core part of user experience design.

However, a growing movement reframes accessibility as a form of care—prioritizing user dignity, autonomy, and contextual needs. This care-led approach recognizes that access is not achieved through conformance alone, but through adaptive systems that honor user variability.

This divergence is reflected in both tooling and developer practices. Automated accessibility testing tools such as axe or WAVE help flag common issues but cover only a fraction of real-world usability barriers, particularly for users with cognitive or learning differences. As a result, teams that rely solely on automation may unknowingly exclude users who fall outside the reach of those heuristics.

Moreover, technical choices—such as the overuse or misuse of ARIA (Accessible Rich Internet Applications) attributes and bloated JavaScript frameworks—can unintentionally introduce new barriers. The 2025 WebAIM Million study found that home pages using ARIA averaged 57 accessibility errors, compared to 27 errors on pages without ARIA \cite{webaim2025}. This highlights the implementation challenges developers face when incorporating ARIA, especially within dynamic, component-based frameworks. More complex sites often attempt to compensate for framework opacity using ARIA, inadvertently layering abstraction over inaccessibility.

This pattern reveals a structural issue: the increasing complexity of front-end ecosystems has raised the cost of accessibility for developers. Even well-meaning teams may struggle to maintain accessible interfaces when faced with shifting frameworks, limited resources, or insufficient guidance. A care-led approach can help mitigate this by shifting from a prescriptive to a flexible model—one that reduces the burden on developers to anticipate all user needs in advance. Instead, by designing systems that allow users to customize their interface according to personal needs, developers can extend access more inclusively and sustainably.

\section{Shared Needs Across Diverse Users}

While neurodivergent users serve as a vital case study for personalized accessibility, the needs they highlight often resonate far beyond any single community. Features that support them—such as improved font clarity, reduced motion, customizable contrast, and scalable layouts-often benefit a broad spectrum of users. These include older adults experiencing sensory decline, individuals encountering situational impairments (e.g., glare, noise, injury), and users navigating cognitive fatigue or stress \cite{w3c21}. This convergence of needs across diverse contexts highlights the value of designing for shared human variability, not just static categories of ability.

Research shows that dyslexia-friendly fonts improve reading speed and comprehension not only for those with dyslexia but also for readers under cognitive strain or distraction \cite{Rello2017}. Similarly, reducing motion and animation aids those with vestibular disorders or sensory sensitivities while also enhancing usability in overstimulating or cluttered environments \cite{w3c21, cogamain}. Enhanced contrast and legibility support users in bright sunlight or visually complex settings, underscoring that temporary impairments can mirror permanent ones in their functional effects \cite{tigwell2018,ntnu}.

At a cognitive level, user-adjustable interfaces align with \textbf{Cognitive Load Theory}, which advocates reducing extraneous cognitive demands to enhance processing \cite{Sweller2011}. Allowing individuals to control visual and interaction settings supports working memory, attention regulation, and mental effort management—benefits not limited to any one diagnostic label \cite{cogamain,bib6}. As Hourcade argues, designing for \textit{universal interactions} means acknowledging that end users differ significantly from developers in needs, constraints, and contexts; effective systems must accommodate this variability from the outset \cite{HourcadeAgency2023}.

The importance of inclusive design also lies in process, not just outcome. Brosnan highlights that neurodiversity encompasses more than just autism or learning disabilities—it reflects a wide range of lived realities. As such, participatory methods must prioritize transferability, credibility, and authenticity by actively including multiple voices and perspectives during design \cite{BrosnanUX2023}. This orientation supports not only functional accessibility but also epistemic justice in the design process itself.

From a business standpoint, these practices yield measurable results. Customization features correlate with higher engagement, lower bounce rates, and increased conversions \cite{Forbes2024, businesscase}. The disability market, encompassing over 1.85 billion people globally with an estimated \$1.9 trillion in disposable income, represents both a moral and economic incentive \cite{IFC2025}. Furthermore, visible commitments to accessibility enhance brand loyalty and public trust \cite{Norman2019}.

Together, this evidence reframes personalization not as an optional add-on or compliance task, but as a cornerstone of effective, inclusive, and future-proof digital design. Embracing shared needs across diverse users means designing with variability in mind—building systems that flex, adapt, and respond to the realities of human difference.

\section{A Case for Autonomy in Interface Customization}
The concept of autonomy is central to creating digital interfaces that truly respect and empower users. Psychological research consistently highlights the importance of dignity, control, and comfort in shaping positive user experiences, particularly for neurodivergent individuals whose needs often fall outside rigid, one-size-fits-all models \cite{DeciRyan2000, BrosnanUX2023}. Providing options that allow users to tailor their experience fosters a sense of agency, reducing frustration and cognitive overload \cite{HourcadeAgency2023}.

Traditional accessibility efforts often emphasize compliance with established standards, positioning users as passive recipients of predetermined accommodations. However, drawing from the field of medicine, Lutfey and Wishner \cite{medicineparallel} distinguish “compliance” from “adherence,” where the latter reflects a model that respects patient autonomy and social context. They argue that while the shift from “compliance” to “adherence” may seem semantic, it represents a profound paradigmatic change—from enforcing rigid rules to enabling personalized engagement.

This analogy is instructive for digital accessibility: compliance offers a baseline of care, ensuring minimum standards, but autonomy, through customizable interfaces, delivers an enhanced, user-centered form of care. By enabling users to define how they interact with digital content, designers acknowledge diverse needs and contexts, encouraging active participation rather than passive acceptance. This approach can better accommodate the variability inherent in neurodivergence and other accessibility dimensions, expanding the scope of solutions beyond the adherence to the checklist to meaningful, personalized support.

In practice, this means moving away from fixed presets or static overlays toward adaptable tools like the \textit{Comfort Mode} toggle, which offers flexible controls without overwhelming users. This strategy aligns with the broader social paradigm shift in healthcare described by Lutfey and Wishner, where understanding user behavior and constraints leads to more effective, respectful, and inclusive outcomes.

Before introducing the practical framework that embodies these principles, we address a common and lingering concern: can accessibility and aesthetics truly coexist?

\subsection{Reconciling Aesthetics and Accessibility}

The persistent notion that accessibility necessitates aesthetic compromise represents one of the most counterproductive barriers to inclusive design. This false dichotomy stems from a fundamental misunderstanding of how visual design and accessibility interact. Rather than being opposing forces, we demonstrate how they can become mutually reinforcing when properly integrated.

Three core misconceptions drive this artificial divide:

First, the assumption that WCAG guidelines inherently limit creative freedom ignores how structured parameters often enhance rather than restrict design quality. Just as poetic forms like sonnets can inspire greater creativity through their constraints, accessibility requirements can focus design efforts on more intentional, effective solutions.

Second, the belief that accessible interfaces must prioritize function over form fails to recognize that aesthetic coherence itself contributes to usability. Research demonstrates that visually harmonious designs \cite{Brath2014} improve both accessibility and user satisfaction when they maintain clear information architecture while employing stylistic elements purposefully.

Third, concerns that personalization fragments brand identity overlook how adaptable design systems can extend rather than dilute visual language. When alternate states are crafted with the same care as default interfaces, they become complementary expressions of brand values rather than compromises.

The Comfort Mode framework resolves these tensions by demonstrating how:
\begin{itemize}
    \item Accessibility requirements can inspire more thoughtful visual hierarchies
    \item Personalization features become natural extensions of brand language
    \item Alternate states receive equal design attention and resources
\end{itemize}

This synthesis provides the foundation for our five-pillar framework, proving that design excellence and accessibility are not merely compatible - they are fundamentally interdependent when approached holistically.
\section{The Comfort Mode Framework}

\subsection{Overview and Rationale}

The \textit{Comfort Mode Framework} proposes a pragmatic and care-centered approach to digital accessibility—one that prioritizes user autonomy, contextual adaptability, and inclusive design. While it is grounded in established research and accessibility principles, it is intentionally crafted to be implementation-flexible across diverse contexts, such as an undergraduate student building her personal portfolio, a startup embedding inclusive design from the ground up, or an established organization piloting accessibility improvements on a limited scale.

\subsection{Naming Justification}

The name \textit{Comfort Mode} is intentionally simple and user-facing, echoing language already familiar to users from settings interfaces, reading platforms, and personalization tools. Rather than adopt clinical or compliance-oriented terminology, the term “Comfort Mode” reflects the framework’s ethos: enabling dignity, ease, and emotional safety through user-controlled customization. This choice aligns with both the cognitive accessibility guidelines from the W3C COGA Task Force~\cite{cogamain} and broader design ethics that emphasize familiarity, emotional resonance, and non-stigmatizing terminology~\cite{Tronto1993, BrosnanUX2023, HourcadeAgency2023}.

\subsection{Applicability Across Contexts}

The framework does not prescribe any single set of technologies, development environments, or implementation pathways. Its utility lies in its adaptability. For a student, it may be a minimal JavaScript-based toggle with a few CSS overrides. For a startup, it could be an integrated design system extension. For larger organizations, it can be gradually introduced as a pilot, with incremental metrics and user feedback informing broader deployment. 

\subsection{Core Pillars of the Framework}

At the heart of the Comfort Mode Framework are five conceptual pillars. These pillars serve as both ethical commitments and practical scaffolding:

\begin{enumerate}
    \item \textbf{User Autonomy}\label{autonomy} \\ 
    Rather than a checklist-driven approach to meet standards post-facto, this pillar centers the user’s agency in customizing their experience. It reframes accessibility not as a constraint but as an opportunity to empower users through self-determined choices~\cite{HourcadeAgency2023, DeciRyan2000, cogamain}.

    \item \textbf{Inclusion Through Personalization}\label{inclusion} \\
    Building on evidence that personalization directly supports cognitive accessibility~\cite{moreno2023simplification,hoehl2016web,waggoner2021inclusive}, this pillar emphasizes modular, user-driven adaptations. These can range from dyslexia-friendly fonts and reduced motion to customizable contrast and content simplification.

    \item \textbf{Design Harmony}\label{harmony} \\
    Rejecting the myth that accessible design must compromise visual appeal, this pillar encourages creative harmony between brand identity and inclusive UX. As Brath and Banissi~\cite{Brath2014} argue, aesthetic “sizzle” can support rather than detract from legibility, attention, and retention.

    \item \textbf{Iterative Co-Design}\label{codesign} \\
    Accessibility should not be “designed for” users, but “designed with” them. This pillar foregrounds participatory design practices~\cite{Muller1993}, iterative user testing, and feedback loops as critical to the success and adoption of comfort-focused features.

    \item \textbf{Designing for Dignity}\label{dignity} \\
    Drawing from care ethics~\cite{Tronto1993} and disability justice~\cite{Shakespeare2006}, this final pillar prioritizes emotional safety, user dignity, and the right to opt in or out of features without being pathologized or labeled.
\end{enumerate}

\subsection{From Principles to Practice}

The Comfort Mode Framework is structured to be modular. While its full implementation supports a robust range of personalizations, it can be adopted incrementally. A minimal implementation may begin with a single toggle—enabling higher contrast, increased font size, and element spacing—while providing an optional accessibility settings page that allows users to vote on or submit desired customizations. This page can serve as a bridge between user needs and future development efforts. More advanced deployments may layer additional user-specific settings, integrating cognitive, situational, and sensory support mechanisms based on evolving user data.

Implementation pathways and interface design models are presented in the following section.

\section{Interface Models and Scalable Implementation}

Building upon the modular structure of the Comfort Mode Framework, this section presents example interface models and technical implementation approaches suited for different levels of adoption. The goal is to illustrate how even minimal implementations can provide meaningful accessibility benefits, while also supporting more advanced, user-driven customization strategies.

\subsection{Simple Toggle}

\begin{figure}[h]
  \centering
  \includegraphics[width=0.95\textwidth]{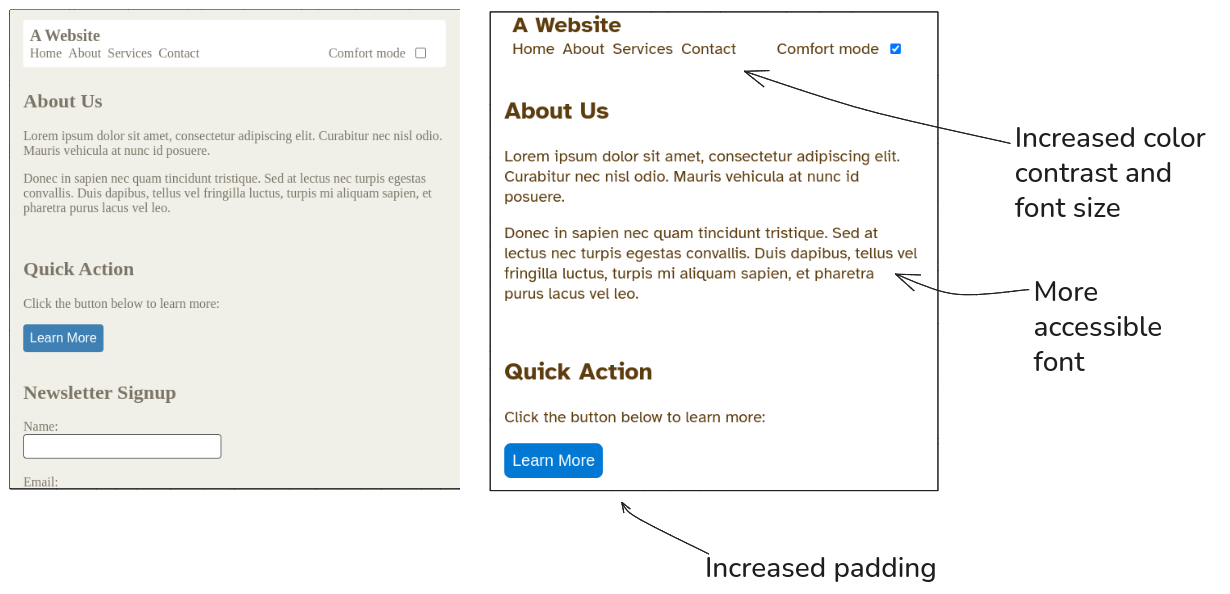}
  \caption{Side-by-side comparison of default and comfort mode interface.}
  \label{fig:comfort-toggle}
\end{figure}

A foundational implementation of Comfort Mode can begin with a single, persistent toggle switch, offering users the option to activate a preset alternative style optimized for readability and comfort. Figure~\ref{fig:comfort-toggle} demonstrates a side-by-side comparison of the default and comfort-enhanced views of a sample webpage.

This basic toggle model implements a small but impactful set of adjustments: increased contrast, larger font size, greater element spacing, and a dyslexia-friendly typeface. These changes align with core principles outlined by the Web Content Accessibility Guidelines (WCAG 2.2)~\cite{w3c22}, particularly in relation to contrast ratios (Success Criterion 1.4.3), text resizing (1.4.4), and visual presentation (1.4.8). These affordances, while general, benefit a broad range of users—including neurodivergent individuals, those with low vision, aging users, and users in temporary or situational impairments.

An important feature of this implementation is that it respects the brand's visual identity. Rather than replacing a brand’s aesthetic, developers can predefine accessible style variables that retain thematic elements—such as color palette, spacing rhythm, and visual tone—while ensuring sufficient accessibility. This approach grants both the platform and the user autonomy: brands can preserve creative intent, and users can personalize their experience based on need.

The toggle can be implemented with minimal code. For example, CSS custom properties can define both the base and comfort mode themes:

\begin{verbatim}
/* Base Theme */
:root {
  --primary-bg: #F0F0E8;
  --primary-text: #7c7567;
  --secondary-bg: #ffffff;
  --secondary-text: #555555;
  --accent-color: #3e80b3;
  --padding: 8px;
  --border-radius: 4px;
  --font-small: 16px;
  --font-large: 20px;
}

/* Comfort Mode Theme */
.comfort {
  --primary-bg: #FFFFFF;
  --primary-text: #5D3C0E;
  --secondary-bg: #ffffff;
  --secondary-text: #555555;
  --accent-color: #0078d4;
  --padding: 10px;
  --border-radius: 8px;
  --font-small: 20px;
  --font-large: 28px;
  --font-family: 'Atkinson Hyperlegible', sans-serif;
}
\end{verbatim}

JavaScript logic to activate the comfort mode is equally lightweight:

\begin{verbatim}
document.getElementById('toggle').addEventListener('change', function() {
  if (this.checked) {
    document.documentElement.classList.add('comfort');
  } else {
    document.documentElement.classList.remove('comfort');
  }
});
\end{verbatim}

This minimal example demonstrates that meaningful customization can be both technically feasible and user-respecting, even at small scales. For complete source code, see Appendix~\ref{appA}.

\subsection{Progressive Disclosure and Personalization}

\begin{figure}[h]
    \centering
    \includegraphics[width=\linewidth]{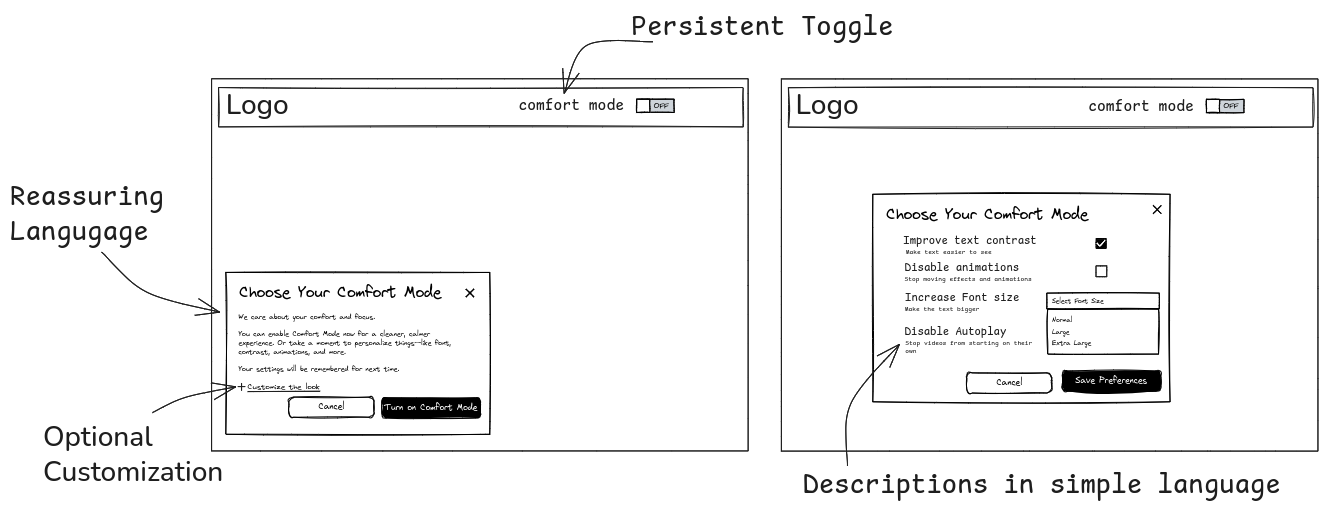}
    \caption{Comfort Mode mockup showing a toggle-based activation and optional layered customization.}
    \label{fig:comfort-mode-mockup}
\end{figure}

The Comfort Mode interface, illustrated in Figure~\ref{fig:comfort-mode-mockup}, demonstrates how accessibility can be approached not just as a compliance checklist, but as an opportunity to offer care and user dignity through thoughtful interaction design. Its layered structure enables users to engage at their own pace—whether by simply toggling a persistent accessibility mode or customizing settings to meet specific needs.

This approach leverages \textbf{progressive disclosure}, a usability strategy that surfaces complexity only when necessary. Such layering reduces cognitive load, particularly for users with executive function differences or anxiety around overwhelming interfaces. At the same time, it avoids the common trap of oversimplification, offering meaningful customization for users who seek more control.

The decision to make Comfort Mode settings persistent—akin to cookie consent banners—signals to users that accessibility is not an afterthought. Rather than hiding options deep within a settings menu, users are reassured that personalization is expected and welcome. This aligns closely with the W3C COGA principle to \textit{“support personalization and allow the user to control and simplify content”}~\cite{w3c21}.

Plain language descriptions such as “Make text bigger” or “Stop moving effects” reinforce WCAG 2.2’s commitment to understandability~\cite{w3c22}. These micro-interactions embody the principle of \textbf{Dignity-Centered Design}, wherein users are not burdened with decoding jargon or navigating inaccessible defaults. Instead, accessibility is framed as a first-class citizen of the interface—stable, obvious, and optional.

The mockup operationalizes two key pillars of the Comfort Mode Framework: \textbf{User Autonomy} and \textbf{Inclusion Through Personalization}. Rather than imposing a one-size-fits-all standard, the interface invites diverse users to define their own comfortable experience. This aligns with the broader movement in inclusive design, which prioritizes flexible systems over fixed accommodations~\cite{BrosnanUX2023, HourcadeAgency2023}.

\subsection{Supporting Infrastructure and Community Feedback}

\begin{figure}[h]
    \centering
    \includegraphics[width=0.75\linewidth]{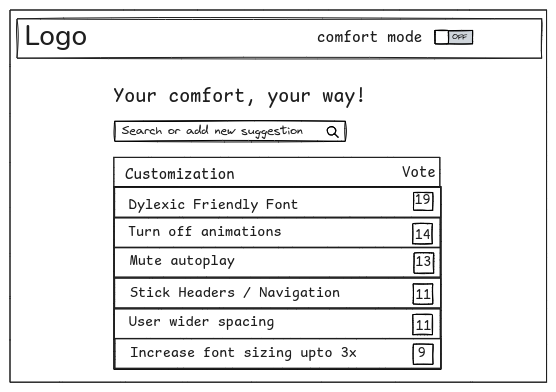}
    \caption{Participatory infrastructure mockup for community-driven personalization of Comfort Mode settings.}
    \label{fig:comfort-mode-feedback}
\end{figure}

Beyond the toggle and customization modal, Figure~\ref{fig:comfort-mode-feedback} represents an optional feedback interface designed to invite user participation in the evolution of accessibility features. This approach treats personalization not as a fixed set of developer assumptions, but as an ongoing dialogue with the community. Users can suggest new Comfort Mode preferences, upvote existing ideas, and surface unmet needs.

This participatory infrastructure shifts accessibility from a prescriptive checklist to a living system of co-authored care. As emphasized in our earlier framework pillars—particularly \textit{Dignity-Centered Design} and \textit{Iterative Co-Design}—users are not merely accommodated; they are welcomed as contributors. This model mirrors the success of community-maintained design systems and preference aggregation platforms in broader open-source and civic tech spaces.

Such a mechanism also helps reduce the guesswork implementers often face when anticipating edge cases or neurodivergent needs. Instead of overengineering a universal solution, developers can enable adaptive pathways based on real user feedback—grounded in context, temporality, and self-identification. As the W3C COGA guidance notes, personalization must be “available when needed” and “relevant to the user’s current task or environment”~\cite{cogamain}.

Depending on their scale and audience, organizations may adopt this system at varying levels of depth. A grassroots project might simply allow voting and suggestions, while a larger brand could embed user surveys, A/B testing, or even real-time cognitive load feedback loops. The goal is not to mandate complexity but to enable care where capacity exists.

This vision points toward promising directions for future work: federated preference storage, community-shared configuration profiles, and adaptive personalization based on session context or individual history. Whether implemented as a lightweight toggle or a deeply modular interface, Comfort Mode can remain technically feasible while rooted in respect, adaptability, and user trust.

\section{Future Work}

\subsection{Research Directions}

While the Comfort Mode Framework presents a conceptual and functional foundation, future research can expand its empirical and theoretical depth. Usability studies across diverse neurodivergent and non-disabled populations can help assess which customization features yield the most meaningful impact, particularly under different cognitive loads or environmental contexts. Longitudinal evaluations may also reveal how user preferences evolve over time, especially when given greater autonomy.

Further inquiry is needed into questions of user identity, federated preference storage, and the ethics of personalization—especially as customization systems intersect with privacy, profiling, and algorithmic adaptation. The framework may also be extended to other modes of interaction beyond visual (e.g., voice interfaces or spatial computing), requiring novel adaptations grounded in multimodal accessibility research.

\subsection{Broader Applications and Tooling}

To support wider adoption, future work can also focus on tooling and open-source infrastructure that lowers the barrier for developers and designers to implement customizable accessibility.

One promising direction is a \textbf{contrast-aware palette generator}—a utility that receives a brand’s color palette and outputs a visually harmonious, WCAG~AAA-compliant variant suitable for Comfort Mode themes. This allows teams to preserve brand identity while meeting the highest readability standards.

Another avenue is an \textbf{open-source preferences hub}, designed as a shared repository of community-vetted customization features. This platform could aggregate suggestions (e.g., dyslexia-friendly fonts, motion-reduction settings, layout toggles), enable voting or prioritization, and maintain public versioning for transparency. While moderation and ethical stewardship are essential to prevent misuse or tokenistic contributions, such a system could accelerate inclusive design across products, especially for smaller teams without dedicated accessibility departments.

Finally, further efforts may explore plug-and-play libraries, CMS integrations, and frontend frameworks that provide Comfort Mode as a drop-in component. Whether through a minimal toggle or a full customization suite, the goal remains the same: to decentralize accessibility responsibility and offer users more agency, not just accommodation.

\section{Conclusion}

Accessibility is not a checkbox—it is a commitment to care. This paper has proposed the Comfort Mode Framework as a modular, dignity-centered approach to inclusive web design, grounded in user autonomy and personalization. By inviting users to tailor their digital experience rather than conform to a single default, we shift accessibility from a reactive fix to a proactive collaboration. Whether implemented through a simple toggle or an evolving customization suite, Comfort Mode affirms that accessibility is not just about reducing complexity for developers—it’s about reducing friction for users. As we move beyond compliance, toward empathy and equity, we call on designers, developers, and institutions alike to build not only for access, but for agency.

\section{Acknowledgments}
My deepest thanks to Mr.Krishna, whose quiet encouragement and patience made space for this work to grow.
\begin{appendices}
\section{ Lightweight JavaScript for Persistent Storage of Preferences}\label{appA}

The following JavaScript snippet demonstrates how the Comfort Mode toggle can be implemented with persistent settings usinglocalStorage, enabling user preferences to remain consistent across sessions without external dependencies.

\begin{verbatim}
    document.addEventListener('DOMContentLoaded', function () {
  const toggle = document.getElementById('toggle');
  const comfortMode = localStorage.getItem('comfortMode') === 'true';
  toggle.checked = comfortMode;
  if (comfortMode) {
    document.documentElement.classList.add('comfort');
  }

  toggle.addEventListener('change', function () {
    if (this.checked) {
      document.documentElement.classList.add('comfort');
    } else {
      document.documentElement.classList.remove('comfort');
    }
    localStorage.setItem('comfortMode', this.checked);
  });
});

\end{verbatim}




\end{appendices}

\section*{Statements and Declarations}
\subsection{Funding Declaration}
No funding was received to assist with the preparation of this manuscript.

\subsection{Author Contribution}
L.A.R. was responsible for all aspects of this manuscript, including conceptualization, methodology, writing the original draft, and review and editing.

\bibliography{sn-bibliography}


\begin{thebibliography}{27}
\ifx \bisbn   \undefined \def \bisbn  #1{ISBN #1}\fi
\ifx \binits  \undefined \def \binits#1{#1}\fi
\ifx \bauthor  \undefined \def \bauthor#1{#1}\fi
\ifx \batitle  \undefined \def \batitle#1{#1}\fi
\ifx \bjtitle  \undefined \def \bjtitle#1{#1}\fi
\ifx \bvolume  \undefined \def \bvolume#1{\textbf{#1}}\fi
\ifx \byear  \undefined \def \byear#1{#1}\fi
\ifx \bissue  \undefined \def \bissue#1{#1}\fi
\ifx \bfpage  \undefined \def \bfpage#1{#1}\fi
\ifx \blpage  \undefined \def \blpage #1{#1}\fi
\ifx \burl  \undefined \def \burl#1{\textsf{#1}}\fi
\ifx \doiurl  \undefined \def \doiurl#1{\url{https://doi.org/#1}}\fi
\ifx \betal  \undefined \def \betal{\textit{et al.}}\fi
\ifx \binstitute  \undefined \def \binstitute#1{#1}\fi
\ifx \binstitutionaled  \undefined \def \binstitutionaled#1{#1}\fi
\ifx \bctitle  \undefined \def \bctitle#1{#1}\fi
\ifx \beditor  \undefined \def \beditor#1{#1}\fi
\ifx \bpublisher  \undefined \def \bpublisher#1{#1}\fi
\ifx \bbtitle  \undefined \def \bbtitle#1{#1}\fi
\ifx \bedition  \undefined \def \bedition#1{#1}\fi
\ifx \bseriesno  \undefined \def \bseriesno#1{#1}\fi
\ifx \blocation  \undefined \def \blocation#1{#1}\fi
\ifx \bsertitle  \undefined \def \bsertitle#1{#1}\fi
\ifx \bsnm \undefined \def \bsnm#1{#1}\fi
\ifx \bsuffix \undefined \def \bsuffix#1{#1}\fi
\ifx \bparticle \undefined \def \bparticle#1{#1}\fi
\ifx \barticle \undefined \def \barticle#1{#1}\fi
\bibcommenthead
\ifx \bconfdate \undefined \def \bconfdate #1{#1}\fi
\ifx \botherref \undefined \def \botherref #1{#1}\fi
\ifx \url \undefined \def \url#1{\textsf{#1}}\fi
\ifx \bchapter \undefined \def \bchapter#1{#1}\fi
\ifx \bbook \undefined \def \bbook#1{#1}\fi
\ifx \bcomment \undefined \def \bcomment#1{#1}\fi
\ifx \oauthor \undefined \def \oauthor#1{#1}\fi
\ifx \citeauthoryear \undefined \def \citeauthoryear#1{#1}\fi
\ifx \endbibitem  \undefined \def \endbibitem {}\fi
\ifx \bconflocation  \undefined \def \bconflocation#1{#1}\fi
\ifx \arxivurl  \undefined \def \arxivurl#1{\textsf{#1}}\fi
\csname PreBibitemsHook\endcsname

\bibitem[\protect\citeauthoryear{Martins and Duarte}{2024}]{martins2024}
\begin{barticle}
\bauthor{\bsnm{Martins}, \binits{B.}},
\bauthor{\bsnm{Duarte}, \binits{C.}}:
\batitle{A large-scale web accessibility analysis considering technology adoption}.
\bjtitle{Universal Access in the Information Society}
\bvolume{23}(\bissue{4}),
\bfpage{1857}--\blpage{1872}
(\byear{2024})
\doiurl{10.1007/s10209-023-01010-0}
\end{barticle}
\endbibitem

\bibitem[\protect\citeauthoryear{{WebAIM}}{2025}]{webaim2025}
\begin{botherref}
\oauthor{\bsnm{{WebAIM}}}:
The WebAIM Million - The 2025 report on the accessibility of the top 1,000,000 home pages.
Accessed: 2025-06-07
(2025).
\url{https://webaim.org/projects/million/}
\end{botherref}
\endbibitem

\bibitem[\protect\citeauthoryear{Tronto}{1993}]{Tronto1993}
\begin{bbook}
\bauthor{\bsnm{Tronto}, \binits{J.C.}}:
\bbtitle{Moral Boundaries: A Political Argument for an Ethic of Care}.
\bpublisher{Routledge},
\blocation{New York}
(\byear{1993})
\end{bbook}
\endbibitem

\bibitem[\protect\citeauthoryear{Lutfey and Wishner}{1999}]{medicineparallel}
\begin{barticle}
\bauthor{\bsnm{Lutfey}, \binits{K.E.}},
\bauthor{\bsnm{Wishner}, \binits{W.J.}}:
\batitle{Beyond ``compliance'' is ``adherence'': Improving the prospect of diabetes care}.
\bjtitle{Diabetes Care}
\bvolume{22}(\bissue{4}),
\bfpage{635}--\blpage{639}
(\byear{1999})
\doiurl{10.2337/diacare.22.4.635}
\end{barticle}
\endbibitem

\bibitem[\protect\citeauthoryear{{W3C Cognitive and Learning Disabilities Accessibility Task Force}}{}]{bib6}
\begin{botherref}
\oauthor{\bsnm{{W3C Cognitive and Learning Disabilities Accessibility Task Force}}}:
Cognitive Accessibility Roadmap and Gap Analysis.
\url{https://w3c.github.io/coga/gap-analysis/}.
W3C Working Draft, 27 March 2025.Accessed 6 June 2025, Work in progress.
\end{botherref}
\endbibitem

\bibitem[\protect\citeauthoryear{Sears et~al.}{2003}]{sears2003}
\begin{bchapter}
\bauthor{\bsnm{Sears}, \binits{A.}},
\bauthor{\bsnm{Lin}, \binits{M.}},
\bauthor{\bsnm{Jacko}, \binits{J.}},
\bauthor{\bsnm{Xiao}, \binits{Y.}}:
\bctitle{When computers fade: Pervasive computing and situationally-induced impairments and disabilities}.
In: \bbtitle{HCI International},
vol. \bseriesno{2},
pp. \bfpage{1298}--\blpage{1302}
(\byear{2003})
\end{bchapter}
\endbibitem

\bibitem[\protect\citeauthoryear{Tigwell et~al.}{2018}]{tigwell2018}
\begin{bchapter}
\bauthor{\bsnm{Tigwell}, \binits{G.W.}},
\bauthor{\bsnm{Flatla}, \binits{D.R.}},
\bauthor{\bsnm{Menzies}, \binits{R.}}:
\bctitle{It's not just the light: understanding the factors causing situational visual impairments during mobile interaction}.
In: \bbtitle{Proceedings of the 10th Nordic Conference on Human-Computer Interaction},
pp. \bfpage{338}--\blpage{351}
(\byear{2018})
\end{bchapter}
\endbibitem

\bibitem[\protect\citeauthoryear{Steinfeld et~al.}{2019}]{steinfeld2019}
\begin{botherref}
\oauthor{\bsnm{Steinfeld}, \binits{A.}},
\oauthor{\bsnm{Zimmerman}, \binits{J.}},
\oauthor{\bsnm{Tomasic}, \binits{A.}}:
Universal design and adaptive interfaces as a strategy for induced disabilities.
arXiv preprint arXiv:1904.06134
(2019)
\end{botherref}
\endbibitem

\bibitem[\protect\citeauthoryear{Moreno et~al.}{2023}]{moreno2023simplification}
\begin{barticle}
\bauthor{\bsnm{Moreno}, \binits{L.}},
\bauthor{\bsnm{Petrie}, \binits{H.}},
\bauthor{\bsnm{Martínez}, \binits{P.}},
\bauthor{\bsnm{Alarcón}, \binits{R.}}:
\batitle{Designing user interfaces for content simplification aimed at people with cognitive impairments}.
\bjtitle{Universal Access in the Information Society}
\bvolume{23},
\bfpage{99}--\blpage{117}
(\byear{2023})
\doiurl{10.1007/s10209-023-00986-z}
\end{barticle}
\endbibitem

\bibitem[\protect\citeauthoryear{Hoehl}{2016}]{hoehl2016web}
\begin{botherref}
\oauthor{\bsnm{Hoehl}, \binits{J.A.}}:
Exploring web simplification for people with cognitive disabilities.
PhD thesis,
University of Colorado at Boulder
(2016).
Doctoral dissertation
\end{botherref}
\endbibitem

\bibitem[\protect\citeauthoryear{Waggoner et~al.}{2021}]{waggoner2021inclusive}
\begin{botherref}
\oauthor{\bsnm{Waggoner}, \binits{T.}},
\oauthor{\bsnm{Jose}, \binits{J.A.}},
\oauthor{\bsnm{Nair}, \binits{A.}},
\oauthor{\bsnm{Manikandan}, \binits{S.}}:
Inclusive design: Accessibility settings for people with cognitive disabilities.
arXiv preprint arXiv:2110.05688
(2021)
\end{botherref}
\endbibitem

\bibitem[\protect\citeauthoryear{W3C}{2018}]{w3c21}
\begin{botherref}
\oauthor{\bsnm{W3C}}:
{Web Content Accessibility Guidelines (WCAG) 2.1}.
\url{https://www.w3.org/TR/WCAG21/}.
Accessed 2025-06-10
(2018)
\end{botherref}
\endbibitem

\bibitem[\protect\citeauthoryear{Rello and Baeza-Yates}{2017}]{Rello2017}
\begin{bchapter}
\bauthor{\bsnm{Rello}, \binits{L.}},
\bauthor{\bsnm{Baeza-Yates}, \binits{R.}}:
\bctitle{Good fonts for dyslexia}.
In: \bbtitle{Proceedings of the 15th International ACM SIGACCESS Conference on Computers and Accessibility},
pp. \bfpage{1}--\blpage{8}
(\byear{2017}).
\doiurl{10.1145/2513383.2513447}
\end{bchapter}
\endbibitem

\bibitem[\protect\citeauthoryear{{W3C Cognitive and Learning Disabilities Accessibility Task Force}}{}]{cogamain}
\begin{botherref}
\oauthor{\bsnm{{W3C Cognitive and Learning Disabilities Accessibility Task Force}}}:
Making Content Usable for People with Cognitive and Learning Disabilities.
\url{https://w3c.github.io/coga/content-usable/}.
W3C Editor's Draft, 27 March 2025. Accessed 6 June 2025, Work in progress.
\end{botherref}
\endbibitem

\bibitem[\protect\citeauthoryear{of~Science and (NTNU)}{2021}]{ntnu}
\begin{botherref}
\oauthor{\bsnm{Science}, \binits{N.U.}},
\oauthor{\bsnm{(NTNU)}, \binits{T.}}:
Universell Design of ICT.
\url{https://www.ntnu.edu/web/universell/universell-design-of-ict}.
Accessed: 2025-06-10
(2021)
\end{botherref}
\endbibitem

\bibitem[\protect\citeauthoryear{Sweller}{2011}]{Sweller2011}
\begin{botherref}
\oauthor{\bsnm{Sweller}, \binits{J.}}:
Cognitive load theory.
The psychology of learning and motivation: Cognition in education,
37--76
(2011)
\doiurl{10.1016/B978-0-12-387691-1.00002-8}
\end{botherref}
\endbibitem

\bibitem[\protect\citeauthoryear{Hourcade and Bullock-Rest}{2011}]{HourcadeAgency2023}
\begin{barticle}
\bauthor{\bsnm{Hourcade}, \binits{J.P.}},
\bauthor{\bsnm{Bullock-Rest}, \binits{N.E.}}:
\batitle{Universal interactions: Challenges and opportunities}.
\bjtitle{Interactions}
\bvolume{18}(\bissue{2}),
\bfpage{76}--\blpage{79}
(\byear{2011})
\end{barticle}
\endbibitem

\bibitem[\protect\citeauthoryear{Brosnan et~al.}{2017}]{BrosnanUX2023}
\begin{barticle}
\bauthor{\bsnm{Brosnan}, \binits{M.}},
\bauthor{\bsnm{Holt}, \binits{S.}},
\bauthor{\bsnm{Yuill}, \binits{N.}},
\bauthor{\bsnm{Good}, \binits{J.}},
\bauthor{\bsnm{Parsons}, \binits{S.}}:
\batitle{Beyond autism and technology: Lessons from neurodiverse populations}.
\bjtitle{Journal of Enabling Technologies}
\bvolume{11}(\bissue{2}),
\bfpage{43}--\blpage{48}
(\byear{2017})
\doiurl{10.1108/JET-02-2017-0007}
\end{barticle}
\endbibitem

\bibitem[\protect\citeauthoryear{{Forbes Technology Council}}{2024}]{Forbes2024}
\begin{botherref}
\oauthor{\bsnm{{Forbes Technology Council}}}:
How Digital Accessibility Can Increase Holiday Sales and Impact Future Growth.
Accessed: 2025-06-10
(2024).
\url{https://www.forbes.com/sites/forbestechcouncil/2024/10/25/how-digital-accessibility-can-increase-holiday-sales-and-impact-future-growth}
\end{botherref}
\endbibitem

\bibitem[\protect\citeauthoryear{Education and Outreach Working~Group}{2023}]{businesscase}
\begin{botherref}
\oauthor{\bsnm{Education}},
\oauthor{\bsnm{Outreach Working~Group}, \binits{W.}}:
The Business Case for Digital Accessibility.
Accessed 2025-06-07
(2023).
\url{https://www.w3.org/WAI/business-case/}
\end{botherref}
\endbibitem

\bibitem[\protect\citeauthoryear{Corporation}{2025}]{IFC2025}
\begin{botherref}
\oauthor{\bsnm{Corporation}, \binits{I.F.}}:
Persons with Disabilities: Economic Inclusion.
Accessed 2025-06-07
(2025).
\url{https://www.ifc.org/en/what-we-do/sector-expertise/gender/economic-inclusion/persons-with-disabilities}
\end{botherref}
\endbibitem

\bibitem[\protect\citeauthoryear{Norman}{2019}]{Norman2019}
\begin{bbook}
\bauthor{\bsnm{Norman}, \binits{D.}}:
\bbtitle{The Design of Everyday Things: Revised and Expanded Edition}.
\bpublisher{Basic Books},
\blocation{New York}
(\byear{2019})
\end{bbook}
\endbibitem

\bibitem[\protect\citeauthoryear{Deci and and}{2000}]{DeciRyan2000}
\begin{barticle}
\bauthor{\bsnm{Deci}, \binits{E.L.}},
\bauthor{\bsnm{and}, \binits{R.M.R.}}:
\batitle{The "what" and "why" of goal pursuits: Human needs and the self-determination of behavior}.
\bjtitle{Psychological Inquiry}
\bvolume{11}(\bissue{4}),
\bfpage{227}--\blpage{268}
(\byear{2000})
\doiurl{10.1207/S15327965PLI1104\_01}
\end{barticle}
\endbibitem

\bibitem[\protect\citeauthoryear{Brath et~al.}{2005}]{Brath2014}
\begin{bchapter}
\bauthor{\bsnm{Brath}, \binits{R.}},
\bauthor{\bsnm{Peters}, \binits{M.}},
\bauthor{\bsnm{Senior}, \binits{R.}}:
\bctitle{Visualization for communication: the importance of aesthetic sizzle}.
In: \bbtitle{Ninth International Conference on Information Visualisation (IV'05)},
pp. \bfpage{724}--\blpage{729}
(\byear{2005}).
\doiurl{10.1109/IV.2005.145}
\end{bchapter}
\endbibitem

\bibitem[\protect\citeauthoryear{Muller and Kuhn}{1993}]{Muller1993}
\begin{barticle}
\bauthor{\bsnm{Muller}, \binits{M.J.}},
\bauthor{\bsnm{Kuhn}, \binits{S.}}:
\batitle{Participatory design}.
\bjtitle{Communications of the ACM}
\bvolume{36}(\bissue{6}),
\bfpage{24}--\blpage{28}
(\byear{1993})
\doiurl{10.1145/153571.255960}
\end{barticle}
\endbibitem

\bibitem[\protect\citeauthoryear{Shakespeare}{2006}]{Shakespeare2006}
\begin{bbook}
\bauthor{\bsnm{Shakespeare}, \binits{T.}}:
\bbtitle{Disability Rights and Wrongs}.
\bpublisher{Routledge},
\blocation{London}
(\byear{2006}).
\doiurl{10.4324/9780203640098}
\end{bbook}
\endbibitem

\bibitem[\protect\citeauthoryear{{World Wide Web Consortium (W3C)}}{2023}]{w3c22}
\begin{botherref}
\oauthor{\bsnm{{World Wide Web Consortium (W3C)}}}:
Web Content Accessibility Guidelines (WCAG) 2.2.
\url{https://www.w3.org/TR/WCAG22/}.
W3C Recommendation, accessed 2025-06-10
(2023)
\end{botherref}
\endbibitem

\end{thebibliography}
\end{document}